\documentclass[12pt,preprint]{aastex}

\usepackage[usenames, dvipsnames]{color}

\shorttitle{CONSTRAINTS ON PLANETESIMAL DISK MASS FROM IAPETUS' CRATERS AND RIDGE}

\shortauthors{RIVERA-VALENTIN ET AL.}

\begin{document}


\title{CONSTRAINTS ON PLANETESIMAL DISK MASS FROM THE CRATERING RECORD AND EQUATORIAL RIDGE ON IAPETUS}


\author{E. G. Rivera-Valentin, A. C. Barr, \& E. J. Lopez Garcia}
\affil{Department of Geological Sciences, Brown University, 324 Brook St, Box 1846, Providence, RI 02912}
\email{rivera-valentin@brown.edu}
\author{M. R. Kirchoff}
\affil{Southwest Research Institute, 1050 Walnut St., Suit 200, Boulder, CO 80302}
\and
\author{P. M. Schenk}
\affil{Lunar and Planetary Institute, 3600 Bay Area Boulevard, Houston, TX 77058}


\begin{abstract}
Iapetus, the outermost regular satellite of Saturn, has a drastic albedo dichotomy and an equatorial circumferential ridge that reaches heights of 20 km and widths of 70 km. This moon is thought to have formed concurrently with Saturn, and so would have experienced an intense bombardment after its formation. The ridge, which has been inferred to be one of the most ancient features on Iapetus' surface, could reasonably be expected to have been eroded by impacts; however, it has retained long continuous sections and a nearly pristine triangular shape with ridge slopes reaching $\sim$ 40$^{\circ}$. We use these observations, along with crater counts on Iapetus' surface, to constrain the total bombardment mass experienced by the satellite since its formation. The ridge morphology and the crater population recorded on Iapetus indicate it received less than 20\% of the bombardment predicted by the classic $\emph{Nice}$ model for early Solar System evolution. Under the recently proposed scenarios of planetsimal-driven migration of the young outer planets including more realistic disk conditions, our results would imply a planetesimal disk mass of $M_{D}\sim12-34M_{\oplus}$.   
\end{abstract}



\section{Introduction}
The outermost regular satellite of Saturn, Iapetus, has many odd features. Among these are a drastic brightness contrast, whereby the trailing hemisphere is ten times brighter than the leading hemisphere \citep{Squyresetal1984, Blackburnetal2011}, a shape consistent with a body spinning every $\sim$16 hours despite its current rotation period of 79 days \citep{Castillo2007}, and a prominent equatorial and nearly circumferential ridge that reaches heights of 20 km and is up to 70 km wide \citep{Porcoetal2005, Giese2008}. This vast ridge system may have originated during a high spin rate period \citep{Porcoetal2005, Castillo2007} or may be the result of debris infall from an ancient impact-generated sub-satellite and/or ring system \citep{Ip2006, Levison2011, Dombardetal2012}. 

The equatorial ridge is considered to be one of the most ancient features on Iapetus because of its crater density, overprinting by basin ejecta, and the general lack of features that pre-date it \citep{Porcoetal2005, Castillo2007, Giese2008}; yet the ridge has retained continuous undisturbed sections that are up to 200 km in length \citep{Porcoetal2005} and nearly pristine peaks \citep{LopezGarcia2014}. Despite its current semi-major axis of $\sim$59 Saturn radii, Iapetus has many more impact basins per unit surface area than the interior mid-sized moons of Saturn \citep{Smithetal1982, Zahnle, Porcoetal2005, KirchoffSchenk}, which would have received more impactors due to gravitational focusing by Saturn. Hence, despite evidence of an intense bombardment across its surface, both the ridge and Iapetus itself have avoided significant disruption during its bombardment history. The observed cratering and geologic record of Iapetus can thus be used to constrain its and the Saturn system bombardment history and provide constraints on dynamical simulations of planeteismal-driven migration during Solar System formation.

The ``classic'' \emph{Nice} model for early Solar System evolution suggests the giant planets formed in a tighter configuration and closer to the sun than their present locations \citep{Gomesetal2005, Tsiganisetal2005}. A Solar System-wide instability was triggered when Jupiter and Saturn crossed their mutual 2:1 mean motion resonance $\sim 700$ Myr after planet formation \citep{Tsiganisetal2005}. This produced a Solar System-wide increase in the rate of impacts from icy and rocky leftovers of planet formation \citep{Gomesetal2005}, similar in magnitude to that inferred from the observed number and clustering of ages of lunar impact basins and rocks, the so-called Late Heavy Bombardment (LHB) \citep{Teraetal1974, Hartmannetal2000}. 

In the classic \emph{Nice} model, the timing of the LHB is strongly dependent on the mass of the disk and the location of the disk edge \citep{Tsiganisetal2005}; however, recent dynamical simulations including the effects of viscous stirring in the disk due to the presence of Pluto-sized objects aim to resolve this sensitivity \citep{Levisonetal2011}. In the  ``\emph{Nice} II" model \citep{Morbidellietal2007, Levisonetal2011}, energy exchanges between the planets and a planetesimal disk containing about a thousand massive Pluto-sized bodies cause an increase in the eccentricity of the inner ice giant, which leads the system to secular resonances that can initiate disk instability for a larger parameter space \citep{Levisonetal2011}, removing much of the sensitivity to initial disk conditions. 

This new scenario predicts a slightly different bombardment history for the outer planet satellites than the classic \emph{Nice} model. Namely, the presence of larger objects in the planetesimal disk increases the eccentricities of objects scattered onto the outer planet satellites. Thus, a smaller number of cometary objects impact the icy moons, but with higher encounter velocities \citep{DonesLevison}. 

Here, we compare results from a Monte Carlo model of impact cratering on Iapetus to its recorded crater history \citep{KirchoffSchenk}, and the state of degradation of its ridge \citep{LopezGarcia2014} to constrain the total bombardment mass experienced by Iapetus. We show that both the global cratering record and the persistence of the ridge suggest a similar bombardment mass of less than 20\% than predicted by the classic \emph{Nice} model. This implies the ridge is ancient though we cannot preclude delaying ridge formation up through the early LHB. Our inferred bombardment mass range is consistent with recently proposed scenarios of planetsimal-driven migration and their predictions about the Saturn system bombardment.


\section{Cratering Record}
\emph{Voyager} observations of the Saturn system suggest the mid-sized icy moons of Saturn were impacted by two populations \citep{Smithetal1981, Smithetal1982}. Population I is characterized by large projectiles that are thought to be early impactors in heliocentric orbits, most probably comets. Population II lacks large impactors and is characterized by a high number density of small craters. This second population is thought to be younger since it dominates relatively young terrain on the inner mid-sized moons \citep{KirchoffSchenk} and may be produced by planetocentric debris launched into orbit by energetic impacts onto the icy satellites \citep{HoredtNeukum, Dobrovolskis2004, Alvarellos2005}. Alternatively, both measured populations may be a result of a single dynamically evolving population of heliocentric impactors \citep{Mintonetal2012}.  

The \emph{Nice} model suggests the projectile source for the Saturn system bombardment was the trans-neptunian disk \citep{Gomesetal2005}, from which the Kuiper Belt formed \citep{Levisonetal2008}. Though the present-day size-frequency distribution (SFD) of the Kuiper Belt is well known (e.g., \citealt{Fraseretal2014}), it is expected that the SFD of a collisionally-interacting population evolves over time due to collisional grinding. \citet{CharnozMorbi}, though, based on inferences of the population of objects in the Scattered Disk and the Oort Cloud, suggest collisional grinding is negligible for the Kuiper Belt. Their work suggest dynamical processes, which are size independent, need to be invoked to explain the mass depletion of the present-day Kuiper belt with respect to expectations for total primitive mass. Thus, the present-day SFD may be similar to the primitive size distribution in the disk \citep{CharnozMorbi}. The size distribution of the objects striking the outer planet satellites may be similar to that of Jupiter's Trojans, which were also populated during the scattering event in the \emph{Nice} model \citep{Morbidellietal2005}, and is similar to the Kuiper Belt \citep{BarrCanup2010, Fraseretal2014}.

Here, we use a Monte Carlo model of global impact cratering \citep{RiveraValentinBarr2014ApJL, RiveraValentinBarr2014EPSL} to simulate the distribution of craters arising on Iapetus. We investigate a variety of possible source populations and total bombardment masses in order to test if the size distribution of the Kuiper Belt can produce a cratering population similar to that measured on Iapetus and provide constraints on the total bombardment mass. 

\subsection{Methods}
We use Monte Carlo methods to select a population of impactors consistent with the SFD of the Kuiper Belt, which generally follows a double power-law in mass ($m$) of the form $dN/dm\propto m^{-q_{1}}$ for impactor diameter $D<D_{B}$ km and $dN/dm\propto m^{-q_{2}}$ for $D>D_{B}$ km. For the cold Kuiper Belt,  $q_{1}=1.6\pm0.1$ and $q_{2}=3.4\pm0.5$ while for the hot Kuiper Belt, $q_{1}=1.3^{+0.2}_{-0.3}$ and $q_{2}=2.4^{+0.2}_{-0.3}$ are the slopes for the small and large objects respectively  \citep{Fraseretal2014}. The break diameter for the cold and hot populations are $D_{B}=140\pm10$ km and $D_{B}=110^{+10}_{-80}$ km. For the Trojan population, $q_{1}=1.6\pm0.2$, $q_{2}=2.7\pm0.3$, and $D_{B}=136\pm8$ km \citep{Fraseretal2014}. Due to the resolution of our 3-D model \citep{RiveraValentinBarr2014ApJL, RiveraValentinBarr2014EPSL}, the smallest impactor considered has $D=10$ km, thus, in contrast to \citet{Charnozetal2009}, we consider a double power law. Because, for all of the investigated source population SFDs, $q_{1}<2$ and $q_{2}>2$, most of the bombardment mass will be delivered by objects with $D\sim D_{B}$ \citep{BarrCanup2010}. Thus our assumption of a double power law will not strongly affect our results.

The total mass of objects striking Iapetus ($M_{B}$) is a free parameter that we vary in small increments between $0.001M_{Nice}\le M_{B}\le2M_{Nice}$, where $M_{Nice}$ is the total bombardment mass predicted by the classic \emph{Nice} model. We extrapolate $M_{Nice}$ for Iapetus based on estimates of the total amount of objects hitting Callisto \citep{BarrCanup2010}, and use the relative impact probabilities from \citet{Zahnle}, which are calculated using dynamical simulations of present-day ecliptic comets \citep{DuncanLevison, LevisonDuncan}. For a 35$M_{\oplus}$ planetesimal disk, Callisto receives $M_{Nice,C}\sim5.4\times10^{20}$~kg of icy material \citep{BarrCanup2010}. The total impacting mass during a classic \emph{Nice} bombardment of Iapetus is 
\begin{equation}
M_{Nice}=M_{Nice,C}\left(\frac{P_{i,I}}{P_{i,C}}\right),
\end{equation}
where $P_{i,I}$ and $P_{i,C}$ are impact probabilities on Iapetus and Callisto respectively. \citet{Zahnle} suggest impact probabilities on Callisto and Iapetus relative to Jupiter of $P_{i,C}=6.1\times10^{-5}$ and $P_{i,I}=1.4\times10^{-6}$. Thus Iapetus receives $\sim2.3\%$ the total bombardment mass on Callisto, which is $M_{Nice}=1.2\times10^{19}$ kg.

The size of a crater produced by a given impactor is estimated using a Pi scaling law \citep{IvanovArt2002}, where an impactor of density $\rho_{i}$ with velocity $v_{i}$ produces a transient crater of diameter, 
\begin{equation}
D_{tc}=1.16\left(\frac{\rho_{i}}{\rho_{s}}\right)D^{0.78}\left(v_{i}\sin\Omega\right)^{0.43}g^{-0.22},
\end{equation}
where $\rho_{s}=1,100$~kg~m$^{-3}$ and $g=0.22$~m~s$^{-2}$ are Iapetus' density and gravity respectively, and $\Omega$ is impact angle, the distribution of which follows $d\Omega=\sin\left(2\Omega\right)$ such that $\Omega=45^{\circ}$ is the most common value \citep{BarrCanup2010}. We assume a nominal impactor density of $\rho_{i}=1,000$~kg~m$^{-3}$ (e.g., \citealt{SternMcKinnon2000}). Impact velocities are simulated following a Rayleigh distribution about a mean value of $\overline{v_{i}} \approx \sqrt{3v_{orb}^2 +v_{\infty}^2 + v_{esc}^2}$ = 8.5~km~s$^{-1}$, where $v_{orb}=3.4$~km~s$^{-1}$ and $v_{esc}=0.6$~km~s$^{-1}$ are Iapetus' Keplerian orbital velocity and escape velocity \citep{Zahnle} respectively, and $v_{\infty}=6$~km~s$^{-1}$ \citep{DonesLevison}.  Our value of $\overline{v_{i}}$ is somewhat higher than reported by \citet{Zahnle} ($\overline{v_{i}}$ = 6.1~km~s$^{-1}$) and \citet{Charnozetal2009} ($\overline{v_{i}}$ = 7.4~km~s$^{-1}$), and arises from increased eccentricities and inclinations kicked up in the disk by Pluto-sized objects resulting in a larger $v_{\infty}$ \citep{DonesLevison}.

Our global Monte Carlo cratering model (see Section 3) has $5\times5$~km pixels such that the smallest simulated crater is $D_{tc}\sim5$ km. The simple-to-complex transition crater diameter ($D_{c}$) on Iapetus is 4.52 km \citep{WhiteSchenk2013}; therefore, all simulated craters are in the complex regime. In this regime, the transient crater diameter is related to the final observed diameter as 
$
D_{fc}=1.34 D_{tc}^{1.11}D_{c}^{-0.11}
$
\citep{McKinnonSchenk1995, SternMcKinnon2000}.

\subsection{Results}
We ran 200 Monte Carlo simulations for each investigated bombardment mass and SFD pair. Because cratering populations typically follow a power law \citep{Melosh1989}, we find the best fit power law to the synthetic cumulative crater frequency for each run following $N(D_{fc})= cD_{fc}^{-q}$, where $N$ is the number of craters with diameters greater than $D_{fc}$ per 10$^{6}$ km$^{2}$, and $c$ and $q$ are fitting parameters (e.g., \citealt{KirchoffSchenk}). The average $c$ and $q$ for the suite of Monte Carlo runs is then recorded for each bombardment mass with standard error of the means to a 95\% confidence. 

For our source population study, we investigate the SFD of the cold and hot Kuiper Belt and the Trojan population within their observed uncertainties \citep{Fraseretal2014}. We find that the synthetic crater populations are more strongly sensitive to $q_{1}$, the small-object distribution, than to $q_{2}$, the large-object distribution. This is because most of the impactors have small diameters, even though most of the total mass hitting Iapetus is delivered by objects with diameters near $D\sim D_{B}$, where the SFD slope break occurs. 

\citet{KirchoffSchenk} measure $q=2.1\pm0.1$ for 10~km $\le D_{fc}\le90$~km. We find that a $q_{1}$ of 1.7, which is within error to the observed value of the cold Kuiper Belt and the Trojan small-object population, produces a synthetic crater population with $q=1.9\pm0.1$, which is within error to the Iapetus measured value. Our inferred small-object slope is in good agreement with the ``Iapetus Scaled Distribution" developed by \citet{Charnozetal2009}, which suggests $q_{1}=1.8$. These results, though, primarily constrain the SFD of impactors striking Iapetus, which is not necessarily equivalent to the original SFD of objects in the planetesimal disk. In our further simulations, we assume the impacting population follows an SFD with parameters of $q_{1}=1.7$ and $q_{2}=3.4$, because these best reproduce the observed crater population.

We then compare the crater density from our synthetic crater populations to the observed crater density on Iapetus to constrain the total mass of objects hitting the satellite. Figure 1 shows the cumulative SFD for Iapetus \citep{KirchoffSchenk} with error following Poisson statistics to 95\% confidence. Additionally, we plot the simulated crater density as a function of bombardment mass. Our results suggest the number density of craters as a function of diameter on Iapetus is best matched by $M_{B}=(0.09\pm0.03)M_{Nice}$, where the error arises from our Monte Carlo methods. Accounting for the error in the measured crater densities extends the possible bombardment mass range to $0.04M_{Nice}\le M_{B}\le0.2M_{Nice}$. Iapetus' cratering, though, may be saturated and thus its crater density may not be fully representative of the total incident bombardment mass; however, we note the inferred mass is valid even for large crater diameters, which will not be affected by saturation.


\section{Ridge Survival}
The continuity of the equatorial ridge on Iapetus and its degradation state can provide an additional constraint on the amount of objects that hit the satellite during its bombardment history. Additionally, when compared with the inferred global total bombardment mass, can also help elucidate ridge formation timing. To remain consistent with observations, the ridge must retain a nearly pristine shape across long continuous sections \citep{Porcoetal2005, LopezGarcia2014}. For the inferred production population's SFD, we find the average impactor radius is $r_{i}\approx 10$~km with a characteristic impact velocity of $v_{i}=8.5$~km~s$^{-1}$, which produces a crater with transient diameter of $D_{tc}\approx94$ km. This covers only a small fraction of the surface of Iapetus, $f=\left(\frac{1}{4}\pi D_{tc}^{2}\right)/\left(4\pi R^{2}\right)$; however, the characteristic number of impacts is large, $N_{i}=\left(R/r_{i}\right)^{3}\left(M_{B}/M\right)$, where $M\approx1.8\times10^{21}$~kg and $R\approx735$~km is the mass and radius of Iapetus respectively, assuming the imapctors have the same density as Iapetus \citep{BarrCanup2010}. Thus, a classic \emph{Nice} model bombardment (i.e, $M_{B}=M_{Nice}$) would excavate a region on Iapetus $N_{crat}=fN_{i}$,
\begin{equation}
N_{crat}=\frac{D_{tc}^{2}}{16}\left(\frac{RM_{B}}{Mr_{i}^{3}}\right),
\end{equation}
which is approximately 2.7 times. This would significantly disrupt the ridge. For $\sim$30\% of the ridge to retain its triangular shape \citep{LopezGarcia2014}, $N_{crat}\le0.7$, which implies $M_{B}\le0.25M_{Nice}$. This simple analytical estimate gives a total bombardment mass that agrees with results from our cratering study; however, this calculation does not account for overlapping impacts or the expected latitudinal distribution of craters, and so is an overestimate. 

\subsection{Methods}
Here we adapt our three-dimensional global Monte Carlo model of impact cratering \citep{RiveraValentinBarr2014ApJL, RiveraValentinBarr2014EPSL} to simulate impact-induced erosion of Iapetus' ridge. Iapetus is modeled as a Cartesian sphere discretized into cubic volume elements 5~km on a side. The sizes, velocities, and impact angles of projectiles hitting the sphere are chosen using the same Monte Carlo methods described in Section 2. Impact locations are chosen randomly in longitude and latitude ($\varphi$) following $d\varphi=\sin\left(2\varphi\right)$ (e.g., \citealt{BarrCanup2010}). We consider the ridge to be a feature centered at the equator that extends $\pm4^{\circ}$ in latitude. The study of ridge topography by \citet{LopezGarcia2014} used images from only a single hemisphere of Iapetus, the dark terrain, where topography is most readily estimated from images; therefore, we analyze the fate of the ridge in only one hemisphere, which is randomly selected for each Monte Carlo run.

For every run, we take consecutive latitudinal slices with width of 5~km ($\sim0.4^{\circ}$ in longitude), at the model resolution, and test if a crater impinges on that slice. If so, we consider that the ridge has lost its triangular shape in that slice -- this is the maximally conservative approach and results in lower bounds on disk mass estimates. For every Monte Carlo simulation, we record the fraction of the ridge that is unmodified by impacts and find the average fraction over the suite of runs for each bombardment mass studied. 

The study of \citet{LopezGarcia2014} also finds that triangular peaks, the dominant observed morphology, composes $31\%\pm11\%$ of each 12$^{\circ}$ longitudinal bin. \citet{Porcoetal2005} find that segments of pristine ridge can be up to 200 km long \citep{Porcoetal2005}. Thus, significant continuous sections of the ridge within the dark terrain have avoided major modification by impacts. We test the continuity of the ridge by dividing the studied hemisphere into 12$^{\circ}$ bins (i.e., $\sim154$~km segments). For every bin, we find the number of non-impacted five kilometer-wide slices. If this value falls within the observed range of 20\% - 42\%, the bin is considered to match observations. For each Monte Carlo run, we find the percentage of successful bins, $P_{success}$. Values of $M_{B}$ that yield $P_{success}$ close to 1 are considered to be consistent with the observed ridge morphology.

\subsection{Results}
Figure 2 shows an example post-bombardment map of Iapetus depicting the number of times a $5\times5$~km area on the surface is excavated by a transient crater cavity for a single impact history with $M_{B}=M_{Nice}$. For this simulation, an area is excavated a maximum of 10 times with an average of $\sim2$ times. We find that $<1\%$ of the ridge survives impact modification. Therefore, as predicted by our analytic model, a classic \emph{Nice} bombardment disrupts the ridge to an extent far greater than observed. 

The percent of the ridge that remains unmodified by impacts as a function of $M_B$ is shown in Figure 3a. Our results suggest that $\sim33\%$ of the ridge remains unmodified by impacts when $M_{B}=(0.22\pm0.02)M_{Nice}$, in agreement with our analytic approach. In Figure 3b, the probability of a successful simulation that produces a ridge consistent with observations is plotted as a function of total bombardment mass. We consider the continuity criterion satisfied for a given $M_{B}$ when the resulting $P_{success}$ is indistinguishable from the peak point, which occurs for $M_{B}=0.2$. Thus, we infer a bombardment mass range of $0.07M_{Nice}\le M_{B}\le0.3M_{Nice}$. Above $0.3M_{Nice}$, too much of the ridge is damaged, while for bombardment masses less than $0.07M_{Nice}$, too much of the ridge survives. 

\section{Discussion}

The global cratering record of Iapetus suggests it suffered a total bombardment mass in the range of $0.04M_{Nice}\le M_{B}\le0.2M_{Nice}$. The continuity and degradation state of the equatorial ridge suggests this feature experienced a total bombardment mass in the range of $0.07M_{Nice}\le M_{B}\le0.3M_{Nice}$. Thus, both the global cratering record of Iapetus and the degradation state of the ridge suggest similar bombardment masses, intersecting at $0.07M_{Nice}\le M_{B}\le0.2M_{Nice}$. Because of the similar implied masses, our results suggest the ridge is one of the most ancient features on Iapetus, in agreement with geological observations \citep{Porcoetal2005, Giese2008, LopezGarcia2014}; however, we can not preclude the possibility of delaying ridge formation to sometime during the LHB, which may occur if the ridge formed by debris infall from a ring \citep{Dombardetal2012}. This is because there exists the possibility Iapetus itself may have experienced $M_{B}\sim0.3M_{Nice}$ while the ridge may have experienced $M_{B}\sim0.07M_{Nice}$. 

Our results suggest Iapetus experienced a smaller bombardment mass than predicted by the classic \emph{Nice} model \citep{Gomesetal2005}. There are several possible explanations. The planetesimal disk mass could have been smaller than the $35M_{\oplus}$ advocated in the classic \emph{Nice} model. Alternatively, Iapetus could have accreted late in Solar System history, missing much of its early bombardment. Indeed, recent work suggest some of Saturn's moons may have been collisionally disrupted during the LHB and re-accreted afterwards \citep{Charnozetal2009, Asphaug}; however, the probability of disrupting Iapetus at its present location during an LHB sourced from a $35M_{\oplus}$ disk is $\le2\%$ \citep{Charnozetal2009}, this value is much smaller considering our inferred $M_{B}$. Indeed, Iapetus is suggested to have formed concurrently with Saturn \citep{Ward1981, CastilloRogezetal2009}. On this basis, we favor the interpretation that, in the classic \emph{Nice} scenario, the planetesimal disk mass is smaller. The initial planetesimal disk mass considered in \citet{Gomesetal2005} of $M_{D}=35M_{\oplus}$ results in a total of $\sim1.2\times10^{19}$~kg of material delivered to Iapetus (see Section 2.1); thus, our results would suggest a planetesimal disk mass of $M_{D}\sim2.5 - 7 M_{\oplus}$. 

In the recent dynamical simulations for Solar System formation, the so-called \emph{Nice} II model \citep{Morbidellietal2007, Levisonetal2011} and ``Jumping Jupiter" \citep{Brasser2009, BatyginBrown2010, Morbidellietal2010, Nesvorny2011, AgnorLin2012, NesvornyMorbi2012, Nesvornyetal2013, Nesvornyetal2014}, a smaller bombardment mass is predicted for the outer satellites \citep{DonesLevison}. This is because the scattered planetesimals are more excited (i.e., highly eccentric and inclined) resulting in higher encounter velocities with the planets. The higher kinetic energy of the planetesimals result in the difficult capture of these bodies via gravitational focussing; thus, less material impacts the planets and their satellites. A planetesimal disk mass of $M_{D}\sim50M_{\oplus}$ leads to a total bombardment mass on the order of $M_{B}\sim0.3M_{Nice}$ on Iapetus \citep{DonesLevison}. Thus, under the Solar System architecture suggested in these dynamical simulations, our results imply a planetesimal disk mass of $M_{D}\sim12 - 34M_{\oplus}$. This is in agreement with the planetesimal disk mass required to reproduce the Jupiter Trojan population, which suggests $M_{D}\sim14-28M_{\oplus}$ \citep{Nesvornyetal2013}, and $M_{D}\sim20M_{\oplus}$ suggested by \citet{NesvornyMorbi2012} in their broad statistical study. Note that we have assumed $v_{\infty}\sim6$~km~s$^{-1}$, consistent with planetesimals with eccentricities and inclinations excited due to gravitational interactions between icy planetesimals and Pluto-sized objects embedded in the disk \citep{Levisonetal2011, DonesLevison}. If other scenarios predict significantly higher or lower $v_{\infty}$, the constraints on disk mass would be adjusted slightly to account for variations in the transient crater diameter arising from faster or slower impactor populations.

The production population on Iapetus is suggested to be similar to the Jupiter Trojans and Kuiper Belt and to be sourced from the trans-Neptunian disk \citep{Gomesetal2005}. Our results suggest the production population incident on Iapetus has a similar SFD to the cold Kuiper Belt and Trojan population. The SFD of the Kuiper Belt is suggested to only have undergone dynamical depletion, which is size independent, and so its present-day distribution should be similar to the primordial SFD \citep{CharnozMorbi}. Hence, our results support dynamical simulations and suggest the Iapetian production population, and by extension that of the Saturn system, is the Kuiper Belt \citep{Gomesetal2005, Mintonetal2012}. Additionally, our inferred SFD and $M_{B}$ are able to reproduce the basin crater density on Iapetus. The measured cumulative basin crater density on Iapetus for $D_{fc}\ge200$~km is $N(200)=1.3\pm0.4$~km$^{-2}$, which is a total of 9 basins \citep{KirchoffSchenk}. The bombardment mass that best replicates this crater density is $M_{B}\sim0.1M_{Nice}$. Assuming average impact characteristics such that $v_{i}=8.5$~km~s$^{-1}$ and $\Omega=45^{\circ}$, and that $\rho_{i}=\rho_{s}$, the total impacting mass required to produce these basins is $7.2\times10^{16}$~kg, which is $\sim6\%$ of the total inferred incident bombardment mass.


\section{Conclusions}
Iapetus, the outer most regular satellite of Saturn, is suggested to have formed concurrently with its parent body \citep{Ward1981, CastilloRogezetal2009}, and have avoided collisional disruption \citep{Charnozetal2009}. Additionally, the unrelaxed nature of its basins relative to the interior moons suggest it did not undergo a significant thermal history \citep{WhiteSchenk2013}. Therefore, Iapetus is expected to have recorded the full outer system bombardment. Among its perplexing features, the two-toned moon of Saturn contains a prominent, semi-continuous equatorial ridge, the geology of which suggests it to be one of the most ancient features on Iapetus \citep{Porcoetal2005, Giese2008}. The ridge, though, has retained long continuous sections and distinct sharp peaks \citep{Porcoetal2005,  LopezGarcia2014} despite the significant impact erosion predicted for this saturnian moon. The recorded cratering record of Iapetus and the geology of its ridge can thus be used to constrain the saturnian bombardment history and provide bounds for Solar System formation models.

Here, we use the recorded crater population of Iapetus \citep{KirchoffSchenk} and the degradation state of its equatorial ridge \citep{LopezGarcia2014}, to investigate the total bombardment mass the satellite experienced. We find the cratering record of Iapetus supports an impactor population with a size distribution consistent with the cold Kuiper Belt and Trojan population. The total mass of objects striking Iapetus is 4\% - 20\% less than predicted by the classic \emph{Nice} model \citep{Tsiganisetal2005, Gomesetal2005}. The well-preserved triangular shape of Iapetus' equatorial ridge and its long continuous segments suggest it experienced only 7\% - 30\% of the classic \emph{Nice} bombardment mass. Therefore, our results suggest the possible bombardment mass experienced by Iapetus lies between $M_{B}\sim0.07-0.2M_{Nice}$, or a total mass $M_B\sim0.84 - 2.4 \times 10^{18}$~kg. 

In the framework of updates to the \emph{Nice} model \citep{Levisonetal2011}, our results suggest a planetesimal disk mass of $M_{D}\sim12-34M_{\oplus}$. This range is in agreement with constraints from the total mass of Jupiter Trojans, which suggest $M_{D}\sim14-28M_{\oplus}$ \citep{Nesvornyetal2013}. Our results are also consistent with constraints on the total bombardment mass arising from limits on impact-induced ice loss, which suggest the LHB delivered less than 10\% of the classic \emph{Nice} bombardment mass \citep{Nimmo}. \citet{BarrCanup2010} and \citet{Barretal2010} suggest $M_D \le 50M_{\oplus}$ in order for both  Callisto and Titan to retain their present-day moment of inertia. The higher $v_{\infty}$ predicted by \citet{DonesLevison} would likely have a small effect on these constraints because Callisto and Titan are deeply embedded in their parent planets' Hill spheres, so $v_{\infty}$ plays less of a role in controlling $\overline{v_{i}}$ than it does for Iapetus.  Therefore, our results not only reproduce the cratering record of Iapetus and the degradation state of its ridge, but are in agreement with a vast array of observations.


\acknowledgments
We thank an anonymous reviewer for valuable comments and L. Dones for helpful conversation. This work was supported by NASA though the Planetary Geology and Geophysics program through grant PG\&G NNX12AI76G. 




\clearpage

\begin{figure}
\plotone{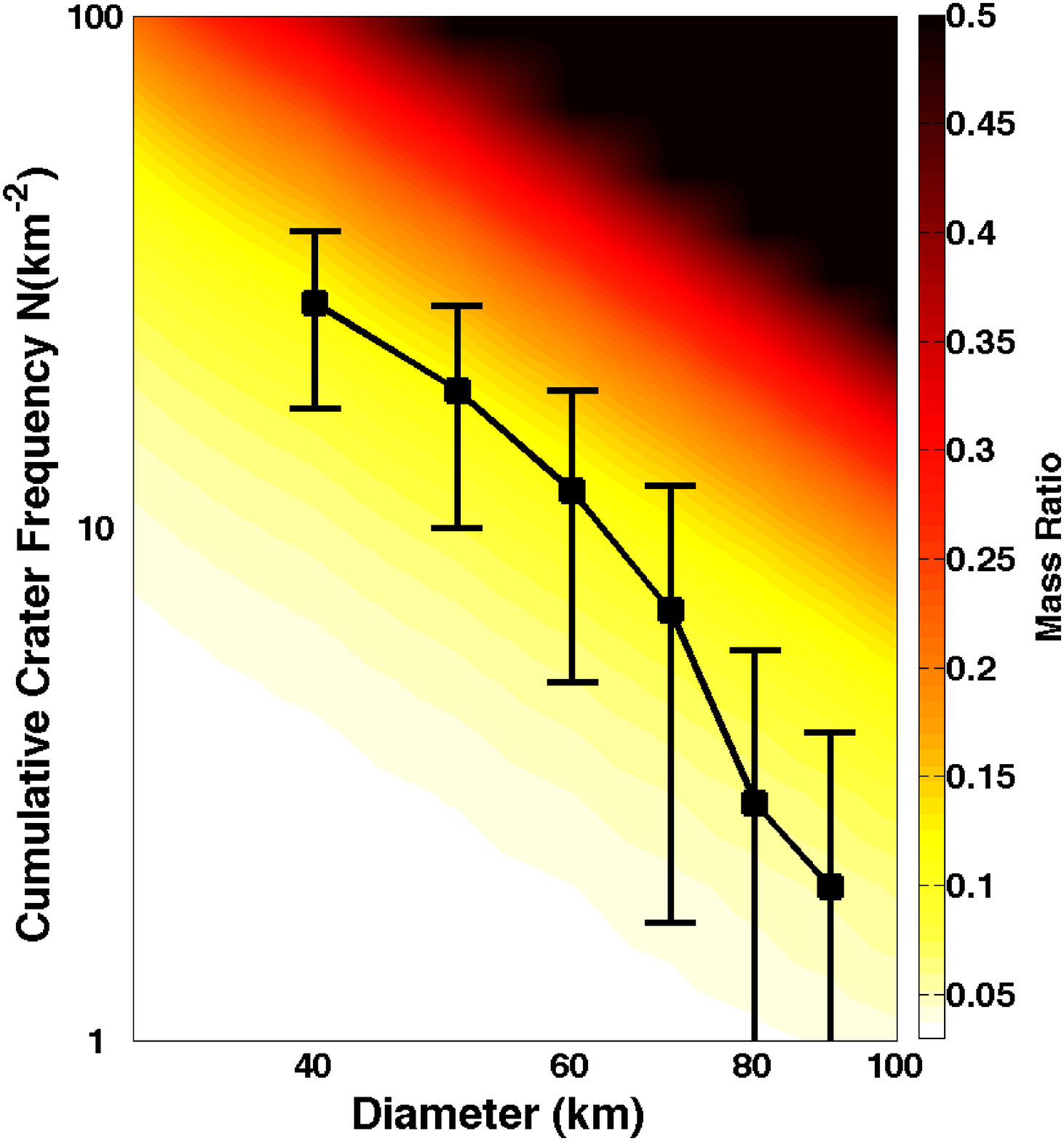}
\caption{Measured cumulative crater size-frequency distribution of the craters on Iapetus (squares). Our model results for the crater density as a function of bombardment mass relative to the classic \emph{Nice} model are depicted by colors. Measured crater densities on Iapetus are best fit by $M_{B}\sim0.1M_{Nice}$ (yellow).}
\end{figure}

\begin{figure}
\plotone{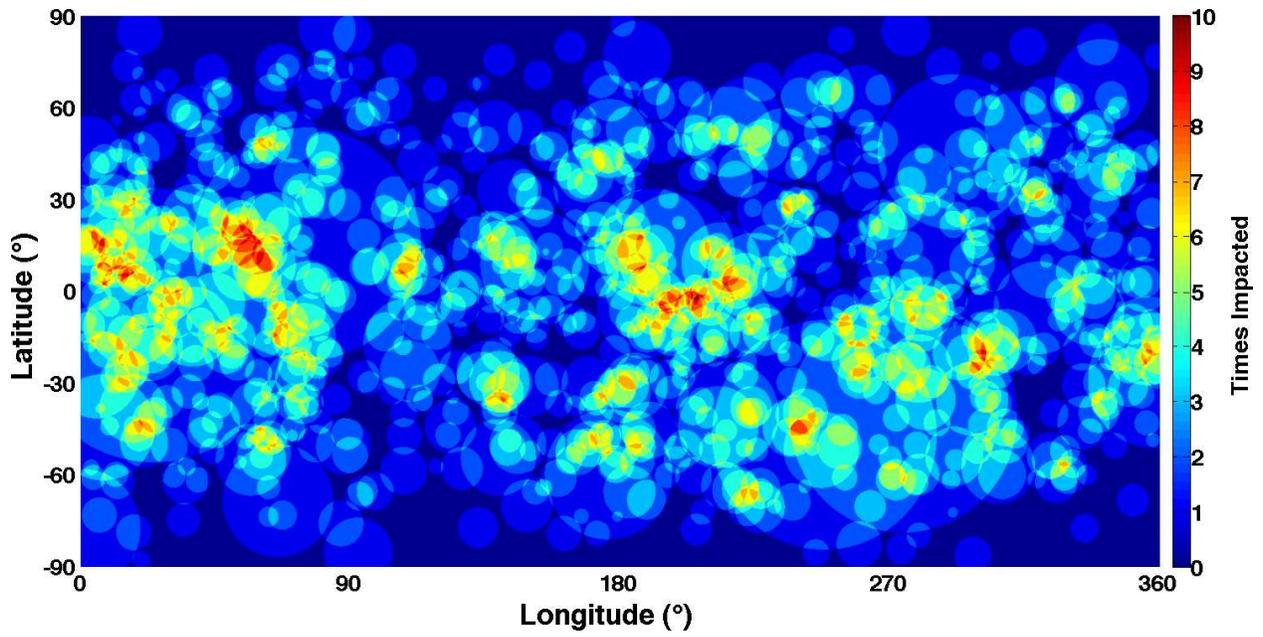}
\caption{Map of the number of times an area is excavated by a transient crater with diameter greater than 5 km for a single Monte Carlo simulation for a full classic \emph{Nice} model bombardment of Iapetus. For this simulation, the maximum number of times an area was excavated is 10, with an average of 2 times.}
\end{figure}

\begin{figure}
\epsscale{0.5}
\plotone{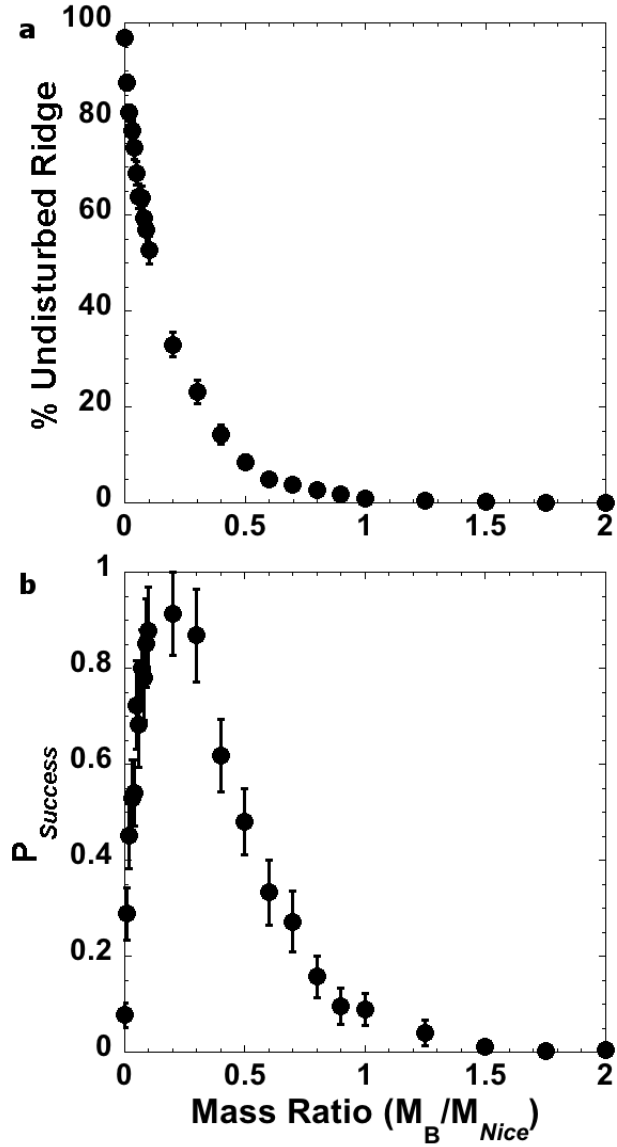}
\caption{Extent of impact modification of the ridge as a function of bombardment mass. a) The percent of the ridge that is unmodified by a transient crater cavity. b) The probability of a successful suite of simulations where the ridge morphology and continuity match observations.}
\end{figure}

\clearpage


\begin{thebibliography}{}
\bibitem[Agnor \& Lin(2012)]{AgnorLin2012} Agnor, C. B. \& Lin, D. N. C. 2012, ApJ, 745, 143

\bibitem[Alvarellos et al.(2005)]{Alvarellos2005} Alvarellos, J. L., Zahnle, K. J., Dobrovolskis, A. R., \& Hamill, P. 2005, Icar, 178, 104. 

\bibitem[Asphaug \& Reufer(2013)]{Asphaug} Asphaug, E., \& Reufer, A. 2013, Icar, 223, 544

\bibitem[Barr \& Canup(2010)]{BarrCanup2010} Barr, A. C. \& Canup, R. M. 2010, NatGe, 3, 164

\bibitem[Barr et al.(2010)]{Barretal2010} Barr, A. C., Citron, R. I., Canup, R. M. 2010, Icar, 209, 858

\bibitem[Batygin \& Brown(2010)]{BatyginBrown2010} Batygin, K. \& Brown, M. E., 2010, ApJ, 716, 1323

\bibitem[Batygin et al.(2012)]{Batygin2012} Batygin, K., Brown, M.~E., Betts, H. 2012, ApJL, 744, L3

\bibitem[Blackburn et al.(2011)]{Blackburnetal2011} Blackburn, D. G., Buratti, B. J., \& Ulrich, R. 2011, Icar, 212, 329

\bibitem[Brasser et al.(2009)]{Brasser2009} Brasser, R., Morbidelli, A., Gomes, R., Tsiganis, K. \& Levison, H. F. 2009, A\&A, 507, 1053

\bibitem[Canup(2010)]{Canup}Canup, R. M. 2010, Natur, 468, 943 

\bibitem[Castillo-Rogez et al.(2007)]{Castillo2007} Castillo-Rogez, J. C., Matson, D. L., Sotin, C. et al. 2007, Icar, 190, 179

\bibitem[Castillo-Rogez et al.(2009)]{CastilloRogezetal2009} Castillo-Rogez, J. C., Johnson, T. V., Lee, M. H. et al. 2009, Icar, 204, 658

\bibitem[Charnoz \& Morbidelli(2007)]{CharnozMorbi} Charnoz, S. \& Morbidelli, A. 2007, Icar, 188, 468

\bibitem[Charnoz et al.(2009)]{Charnozetal2009} Charnoz, S., Morbidelli, A., Dones, L., Salmon, J. 2009, Icar, 199, 413

\bibitem[Charnoz et al.(2011)]{Charnozetal2011} Charnoz, S., Crida, A., Castillo-Rogez, J. C. et al. 2011, Icar, 216, 535

\bibitem[Dobrovolskis \& Lissauer(2004)]{Dobrovolskis2004} Dobrovolskis, A. R. \& Lissauer, J. J. 2004, Icar, 169, 462

\bibitem[Dombard et al.(2012)]{Dombardetal2012} Dombard, A. J., Cheng, A. F., McKinnon, W. B., \& Kay, J. P. 2012, JGRE, 117, E03002

\bibitem[Dones \& Levison(2013)]{DonesLevison} Dones, L. \& Levison, H. 2013, LPSC, 2772

\bibitem[Duncan \& Levison(1997)]{DuncanLevison} Duncan, M. J. \& Levison, H. F. 1997, Sci, 276, 1670

\bibitem[Fraser et al.(2014)]{Fraseretal2014} Fraser, W. C., Brown, M. E., Morbidelli, A., Parker, A., \& Batygin, K. 2014, ApJ, 782, 100

\bibitem[Giese et al.(2008)]{Giese2008} Giese, B., Denk, T., Neukum, G. et al. 2008, Icar, 193, 359
 
\bibitem[Gomes et al.(2005)]{Gomesetal2005} Gomes, R., Levison, H. F., Tsiganis, K., \&  Morbidelli, A. 2005, Natur, 435, 466

\bibitem[Hartmann et al.(2000)]{Hartmannetal2000} Hartmann, W. K., Ryder, G., Dones, L., et al. 2000, in Origin of the Earth and Moon, ed. R. M. Canup et al. (Tuscon, AZ:Univ. of Arizona Press), 493

\bibitem[Horedt \& Neukum(1984)]{HoredtNeukum} Horedt, G. P. \& Neukum, G. 1984, JGRE, 89, 10405
	
\bibitem[Ip(2006)]{Ip2006} Ip, W. H. 2006, GeoRL, 33, L16203

\bibitem[Ivanov \& Artemieva(2002)]{IvanovArt2002} Ivanov, B. A, \& Artemieva, N. A. 2002, GSASP, 356, 619
	
\bibitem[Kirchoff \& Schenk(2010)]{KirchoffSchenk} Kirchoff, M. R. \& Schenk, P. M. 2010, Icar, 206, 485 

\bibitem[Levision \& Duncan(1997)]{LevisonDuncan} Levison, H. F. \& Duncan, M. J. 1997, Icar, 127, 13

\bibitem[Levison et al.(2001)]{Levisonetal2001} Levison, H. F., Dones, L., Chapman, C. R. et al. 2001, Icar, 151, 286

\bibitem[Levison et al.(2008)]{Levisonetal2008} Levison, H. F., Morbidelli, A., VanLaerhoven, C., Gomes, R., \& Tsiganis, K. 2008, Icar, 196, 258 

\bibitem[Levison et al.(2011a)]{Levison2011} Levison, H. F., Walsh, K. Barr, A., \& Dones, L. 2011, Icar, 214, 773

\bibitem[Levison et al.(2011b)]{Levisonetal2011} Levison, H. F., Morbidelli, A., Tsiganis, K., Nesvorny, D., \& Gomes, R. 2011, AJ, 142, 152

\bibitem[Lopez Garcia et al.(2014)]{LopezGarcia2014} Lopez Garcia, E. J., Rivera-Valentin, E. G., Schenk, P. M., Hammond, N. P., \& Barr, A. C. 2014, Icar, 237, 419


\bibitem[McKinnon \& Schenk(1995)]{McKinnonSchenk1995} McKinnon, W. B., \& Schenk, P. M. 1995, GeoRL, 22, 1829

\bibitem[Melosh (1989)]{Melosh1989} Melosh, H. J. 1989, in Impact Cratering: A Geologic Process (New York:Oxford Univ. Press)

\bibitem[Minton et al.(2012)]{Mintonetal2012} Minton, D. A., Richardson, J. E., Thomas, P., Kirchoff, M., \& Schwamb, M. E. 2012, LPSC, 2669 

\bibitem[Morbidelli et al.(2001)]{Morbidellietal2001} Morbidelli, A., Petit, J. M., Gladman, B. \& Chambers, J. 2001, M\&PS, 36, 371

\bibitem[Morbidelli et al.(2005)]{Morbidellietal2005} Morbidelli, A., Levison, H. F., Tsiganis, K., \& Gomes, R. 2005, Natur, 435, 462

\bibitem[Morbidelli et al.(2007)]{Morbidellietal2007} Morbidelli, A., Tsiganis, K., Crida, A., Levison, H. F., \& Gomes, R. 2007, AJ, 134, 1790

\bibitem[Morbidelli et al.(2010)]{Morbidellietal2010} Morbidelli, A., Chambers, J., Lunine, J. I., et al. 2010, M\&PS, 35, 1309

\bibitem[Nesvorn\'y (2011)]{Nesvorny2011} Nesvorn\'y, D. 2011, ApJL, 743, L22

\bibitem[Nesvorn\'y \& Morbidelli(2012)]{NesvornyMorbi2012} Nesvorn\'y, D. \& Morbidelli, A. 2012, AJ, 144, 117

\bibitem[Nesvorn\'y et al.(2013)]{Nesvornyetal2013} Nesvorn\'y, D., Vokrouhlicky, D., \& Morbidelli, A. 2013, ApJ, 768, 45

\bibitem[Nesvorn\'y et al.(2014)]{Nesvornyetal2014} Nesvorn\'y, D., Vokrouhlick\'y, D., \& Deienno, R. 2014, ApJ, 784, 22

\bibitem[Nimmo \& Korycansky(2012)]{Nimmo} Nimmo, F., \& Korycansky, D. G. 2012, Icar, 219, 508

\bibitem[Porco et al.(2005)]{Porcoetal2005} Porco, C. C., Baker, E., Barbara, J. et al. 2005, Sci, 307, 1237

\bibitem[Rivera-Valentin \& Barr(2014a)]{RiveraValentinBarr2014ApJL} Rivera-Valentin, E. G. \& Barr, A. C. 2014, ApJL, 782, L8

\bibitem[Rivera-Valentin \& Barr(2014b)]{RiveraValentinBarr2014EPSL} Rivera-Valentin, E. G. \& Barr, A. C. 2014, E\&PSL, 391, 234
	
\bibitem[Robuchon et al.(2010)]{Robuchonetal2010} Robuchon, G., Choblet, G., Tobie, G. et al. 2010, Icar, 207, 959

\bibitem[Smith et al.(1981)]{Smithetal1981} Smith, B. A., Soderbloom, R., Beebe, R. et al. 1981, Sci, 212, 163

\bibitem[Smith et al.(1982)]{Smithetal1982} Smith, B. A., Soderblom, L., Batson, R. et al. 1982, Sci, 215, 504

\bibitem[Stern \& McKinnon(2000)]{SternMcKinnon2000} Stern, S. A. \& McKinnon, W. B. 2000, AJ, 119, 945

\bibitem[Squyres et al.(1984)]{Squyresetal1984} Squyres, S. W., Buratti, B., Veverka, J., \& Sagan, C. 1984, Icar, 59, 426

\bibitem[Tera et al.(1974)]{Teraetal1974} Tera, F., Papanastassiou, D. A., \& Wasserburg, G. J. 1974, E\&PSL, 22, 1	

\bibitem[Tsiganis et al.(2005)]{Tsiganisetal2005} Tsiganis, K., Gomes, R., Moribidelli, A., \& Levison, H. 2005, Natur, 435, 459

\bibitem[Ward (1981)]{Ward1981} Ward, W. R. 1981, Icar, 46, 97

\bibitem[White et al.(2013)]{WhiteSchenk2013} White, O. L., Schenk, P. M., \& Dombard, A. J. 2013, Icar, 223, 699

\bibitem[Zahnle et al.(2003)]{Zahnle} Zahnle, K., Schenk, P., Levison, H., \& Dones, L. 2003, Icar, 163, 263

\end{thebibliography}
\end{document}